\documentclass[12pt]{iopart}
\usepackage{bm}
\usepackage{graphicx}
\usepackage{amssymb}
\usepackage{hyperref}
\usepackage{color}

\def\wF{{\omega_F}}

\def\Re{{\rm   Re}}

\newcommand{\la}{\langle}
\newcommand{\ra}{\rangle}
\newcommand{\lal}{\langle\langle}
\newcommand{\rar}{\rangle\rangle}
\newcommand{\up}{|\uparrow\rangle}
\newcommand{\down}{|\downarrow\rangle}

\def\hb0{h_{\rm   b}^{(0)}}
\def\p12{p_{12}({\bf q},t)}

\begin{document}
\title {Fluctuation spectra of weakly driven nonlinear systems}

\author{Yaxing Zhang$^{1}$, Yukihiro Tadokoro$^{2}$, and M. I. Dykman$^{1}$}
\address{$^{1}$Department of Physics and Astronomy, Michigan State University, East Lansing, MI 48824, USA}
\address{$^2$Toyota Central R\&D Labs., Inc., Nagakute, Aichi 480-1192, Japan}

\date{\today}
\begin{abstract}
We show that in periodically driven systems, along with the delta-peak at the driving frequency, the spectral density of fluctuations displays extra features. These can be peaks or dips with height quadratic in the driving amplitude, for weak driving. For systems where inertial effects can be disregarded, the peaks/dips are generally located at zero frequency and at the driving frequency. 
The shape and intensity of the spectra very sensitively depend on the parameters of the system dynamics. To illustrate this sensitivity and the generality of the effect, we study three types of systems: an overdamped Brownian particle (e.g., an optically trapped particle), a two-state system that switches between the states at random, and a noisy threshold detector. The analytical results are in excellent agreement with numerical simulations. 
\end{abstract}

\section{Introduction}

Fluctuation spectra and spectra of response to periodic driving are major tools of characterizing physical systems. The spectra are conventionally used to find system frequencies and relaxation rates and to characterize fluctuations in the system. For example, optical absorption spectra give the transition frequencies of atomic systems and the lifetimes of the excited states, and the spectrum of spontaneous radiation is a well-known example of the fluctuation (power) spectrum \cite{Mandel1995}. In macroscopic systems the spectra are often complicated by the effects of inhomogeneous broadening. Recent progress in nanoscience has made it possible to study the spectra of individual dynamical systems. A well-known example is provided by optically trapped Brownian particles and biomolecules \cite{Ashkin1997,Greenleaf2007}, where the power spectra are a major tool for characterizing the motion in the trap \cite{Berg-Sorensen2004,Mas2013}. Spectra of various types of individually accessible mesoscopic systems are studied nowadays in optics \cite{Walls2008,Carmichael2008}, nanomechanics and circuit quantum electrodynamics, cf.~\cite{Dykman2012b}, biophysics, cf.~\cite{Jaramillo1998,Bialek2012},  and many other areas; the technique based on spectral measurements has found various applications, photonic force microscopy being a recent example, see Ref.~\cite{Marago2013}. 

A familiar effect of weak periodic driving is forced vibrations of the system. When ensemble-averaged, they are also periodic and occur at the driving frequency $\omega_F$. They lead to a $\delta$-shape peak at frequency $\omega_F$ in the system power spectrum. However, the driving also modifies the power spectrum away from $\omega_F$. A textbook example is inelastic light scattering and resonance fluorescence. In the both cases, the system driven by a periodic electromagnetic field emits radiation at frequencies that differ from the driving frequency \cite{Heitler2010}. This radiation is one of the major sources of information about the system in optical experiments.

In this paper we study  the spectra of periodically driven nonlinear systems. We show that, in the presence of noise, along with the $\delta$-shape peak at the driving frequency $\omega_F$, these spectra display a characteristic structure. We are interested in the regime of relatively weak driving, where the driving-induced change of the power spectrum is quadratic in the amplitude of the driving, as in inelastic light scattering. 

In view of the significant interest in the power spectra of systems optically trapped in fluids, we consider systems where inertial effects play no role. In the absence of driving the power spectra of such systems usually have a peak at zero frequency. In particular it is this peak that is used to characterize the dynamics of optically trapped particles.

For a linear system, like a Brownian particle in a harmonic trap, the $\delta$-shape peak at $\omega_F$ is the only effect of the driving on the power spectrum. This is because motion of such a system is a linear superposition of forced vibrations at $\omega_F$ and fluctuations in the absence of driving. The amplitude and phase of the forced vibrations depend on the parameters of the system and determine the standard linear susceptibility  \cite{LL_statphys1}. In nonlinear systems forced vibrations become random, because the parameters of the system are fluctuating. The power spectrum of such random vibrations is no longer just a $\delta$-shape peak (although the $\delta$-shape peak is necessarily present). The driving-induced spectral features away  from $\omega_F$ result from mixing of fluctuations and forced vibrations in a nonlinear system. We note the close relation of these features to inelastic light scattering, see \ref{App_susceptibility}.

\subsection{Qualitative picture}

The idea of the driving-induced change of the power spectrum can be gained by looking at a Brownian particle fluctuating in a confining potential, a typical situation for optical trapping. The motion of the particle, after proper rescaling of time and particle coordinate $q$, is described by the Langevin equation \cite{Langevin1908}
\begin{equation}
\label{eq:Brownian}
\dot q = - U'(q) + f(t), \qquad U'(q)\equiv dU/dq,
\end{equation}
where $U(q)$ is the scaled potential and $f(t)$ is thermal noise. If potential $U(q)$ is parabolic and the system is additionally driven by a force $F\cos\omega_F t$, forced vibrations  are described by the textbook expression
\begin{equation}
\label{eq:trivial_susceptibility}
\langle q(t)\rangle =  \frac{1}{2}F\chi(\omega_F)\exp(-i\omega_Ft) + {\rm c.c.}, \qquad \chi(\omega)=[U''(q_{\rm eq})-i\omega]^{-1},
\end{equation}
where $q_{\rm eq}$ is the equilibrium position [the minimum of $U(q)$] and $\chi(\omega)$ is the susceptibility. 

\begin{figure}[h]
\begin{center}
\includegraphics[width=6cm]{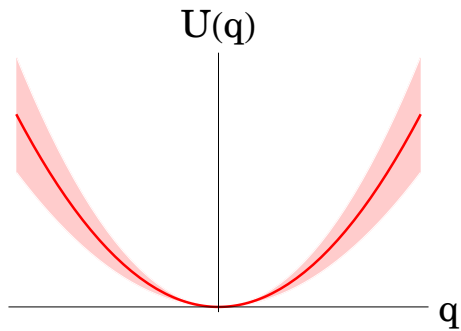}
\caption{Sketch of a potential of a nonlinear system near the potential minimum. Because of the interplay of nonlinearity and fluctuations, the curvature of the potential fluctuates. These fluctuations are shown as the smearing of the solid line, which represents the potential in the absence of fluctuations.}
\label{fig:fluct_potential}
\end{center}
\end{figure}

For a nonlinear system the potential $U(q)$ is nonparabolic. Because of thermal fluctuations, the local curvature of the potential $U''(q)$ is fluctuating. Intuitively, one can think of the effect of thermal fluctuations on forced vibrations as if $U''(q_{\rm eq})$ in Eq.~(\ref{eq:trivial_susceptibility}) for the susceptibility were replaced by a fluctuating curvature, see Fig.~\ref{fig:fluct_potential}.  If the driving frequency $\omega_F$ largely exceeds the reciprocal correlation time of the fluctuations $t_c^{-1}$, the fluctuations would lead to the onset of a structure in the power spectrum near frequency $\omega_F$ with typical  width $t_c^{-1}$. The quantity $t_c^{-1}$ also gives the typical width of the peak in the power spectrum at zero frequency in the absence of driving [for a linear system, $t_c^{-1}=U''(q_{\rm eq})$].

 Another effect of the interplay of  driving, nonlinearity, and fluctuations can be understood by noticing that the periodic force causes a periodic change in the system coordinate. For a nonlinear system, roughly speaking, this leads to a periodic modulation of the local curvature, and thus of $t_c^{-1}$. Since $t_c^{-1}$ determines the shape of the zero-frequency peak in the power spectrum, such modulation causes a change of this peak proportional to $F^2$, to the lowest order in $F$. 

Even from the above simplistic description it is clear that the driving-induced change of the spectrum is sensitive to the parameters of the system and the noise and to the nonlinearity mechanisms. Explicit examples given below demonstrate this sensitivity and suggest that the effects we discuss can be used for characterizing a system beyond the conventional linear analysis. After formulating how the power spectrum can be evaluated in Section~\ref{sec:general}, we demonstrate the effects of the interplay of driving and fluctuations for three very different types of nonlinear systems: an overdamped Brownian particle (Section~\ref{sec:Brownian_particle}), a system that switches at random between coexisting stable states (Section~\ref{sec:two_state}), and a threshold detector (Section~\ref{sec:threshold}). All these systems are of broad interest, and all of them display a significant driving-induced change of the power spectrum.

\section{General formulation}
\label{sec:general}

We consider fluctuating systems driven by a periodic force $F\cos(\wF t)$ and assume that fluctuations are induced by a stationary noise, like in the case of an optically trapped Brownian particle, for example. After a transient time such system reaches a stationary state. The stationary probability distribution of the system with respect to its dynamical variable $q$,  $\rho_{\rm   st}(q,t)$, is periodic in time $t$ with the driving period $\tau_F=2\pi/\omega_F$. The two-time correlation function $\la q(t_1)q(t_2)\ra$ [$\la\cdot\ra$ implies ensemble averaging] is a function of $t_1-t_2$ and a periodic function of $t_2$ with period $\tau_F$. The power spectrum usually measured in experiment is of the form
\begin{eqnarray}
\label{eq:power_spectrum}
\Phi(\omega)&=2 \Re \int_0^\infty dt e^{i\omega t}\lal q(t+t')q(t')\rar,\nonumber\\
\lal q(t&+t')q(t')\rar = \frac{1}{\tau_F}\int_0^{\tau_F} dt'\la q(t+t')q(t')\ra.
\end{eqnarray}

The correlation function in Eq.~(\ref{eq:power_spectrum}) can be expressed in terms of $\rho_{\rm   st}(q,t)$ and the transition probability density $\rho(q_1,t_1|q_2,t_2)$ that the system that was at position $q_2$ at time $t_2$ is at $q_1$ at time $t_1\geq t_2$,
\begin{eqnarray}
\label{eq:corr}
\la q(t_1)q(t_2)\ra = \int dq_1dq_2 \,q_1q_2\,\rho(q_1,t_1|q_2,t_2)\rho_{\rm   st}(q_2,t_2).
\end{eqnarray}
For weak driving, function $\rho_{\rm st}(q,t_2)$ can be expanded in a series in $F\exp(\pm i\omega_Ft_2)$ with time-independent coefficients, whereas $\rho(q_1,t_1|q_2,t_2)$ can be expanded in $F\exp(\pm i\omega_Ft_2)$ with coefficients that depend on $t_1-t_2$. Therefore the power spectrum (\ref{eq:power_spectrum}) does not have terms linear in $F$.  To the second order in $F$ for $\omega\geq 0$ we have
\begin{equation}
\label{eq:define_Phi_F}
\Phi(\omega)=\Phi_0(\omega)+ \frac{\pi}{2}F^2|\chi(\omega_F)|^2\delta(\omega-\omega_F) + F^2\Phi_F(\omega).
\end{equation}
The term $\Phi_0(\omega)$ describes the power spectrum of the system in the absence of driving. The term $\propto\delta(\omega-\omega_F)$ describes the conventional linear response, cf. Eq.~(\ref{eq:trivial_susceptibility}). However, the expression for the susceptibility $\chi(\omega)$ in nonlinear systems is far more complicated than  Eq.~(\ref{eq:trivial_susceptibility}); generally, the susceptibility is determined by the linear in $F$ term in $\rho_{\rm st} (q,t)$. In the optical language, the term $\propto \delta(\omega - \omega_F)$ in (\ref{eq:define_Phi_F}) corresponds to elastic scattering of the field $F\cos\omega_Ft$ by the system. 

Of primary interest to us is the term $\Phi_F(\omega)$. This term is often disregarded in the analysis of the power spectra  of driven systems, while the major emphasis is placed on the $\delta$-function in Eq.~(\ref{eq:define_Phi_F}). Function $\Phi_F(\omega)$ describes the interplay of fluctuations and driving in a nonlinear system beyond the trivial linear response. In the considered lowest-order approximation in the driving amplitude, $\Phi_F$ does not contain a $\delta$-peak at $2\omega_F$. However, it may contain a $\delta$-peak at $\omega=0$, which corresponds to the static driving-induced shift of the average position of the system. In what follows we do not consider this peak, as the static equilibrium position can be measured independently.

Function $\Phi_F$ can be found from Eq.~(\ref{eq:corr}) by calculating the transition probability density and the stationary probability distribution. This can be done for Markov systems numerically and also, in the case of weak noise, analytically, see Secs.~\ref{sec:Brownian_particle} and \ref{sec:two_state}. Alternatively, function $\Phi_F$ can be related to fluctuations of linear and nonlinear response of the system and expressed in terms of the fluctuating linear and nonlinear susceptibility, see \ref{App_susceptibility}. We emphasize that the nonlinear response has to be taken into account when fluctuations are considered even though we are not interested in the behavior of the power spectrum near $2\omega_F$ or higher overtones or subharmonics of $\omega_F$.

\section{Power spectrum of a driven Brownian particle}
\label{sec:Brownian_particle}

A simple example of a system where $\Phi_F(\omega)$ displays a nontrivial behavior is a periodically driven overdamped Brownian particle in a nonlinear confining potential $U(q)$, see Eq.~(\ref{eq:Brownian}). This model immediately relates to many experiments on optically trapped particles and molecules. We will assume that thermal noise $f(t)$ is white and Gaussian and that it is not strong, so that it suffices to keep the lowest-order nontrivial terms in the potential,
\begin{eqnarray}
\label{eq:nonlinear_potential}
U(q)=\frac{1}{2}\kappa q^2 + \frac{1}{3}\beta q^3 + \frac{1}{4}\gamma q^4 + \ldots, \qquad \langle f(t)f(t')\rangle = 2D\delta(t-t'),
\end{eqnarray}
where $D\propto k_BT$ is the noise intensity. In the absence of driving the stationary probability distribution is of the Boltzmann form, $\rho_{\rm st}^{(0)}\propto \exp[-U(q)/D]$.

For small $D$ and weak driving force equation of motion $\dot q=-U'(q)+ f(t) + F\cos\omega_Ft$ can be solved directly by perturbation theory in the noise $f(t)$ and in $F$, as indicated in \ref{App_susceptibility}. Here we develop a different method, which is particularly convenient if one wants to go to high orders of the perturbation theory in $D$ and $F$. 

\subsection{Method of Moments}
\label{subsec:method_of_moments}

Systems in which fluctuations are induced by white noise can be studied using the Fokker-Planck equation 
\begin{equation}
\label{eq:FP_equation}
\partial_t\rho = -\partial_q\left\{\left[-U'(q) + F\cos\omega_Ft\right]\rho\right\} + D\partial^2_q\rho.
\end{equation}
This equation can be solved numerically. A convenient analytical approach is based on the method of moments, which are defined as 
\begin{equation}
\label{eq:moments_defined}
M_n(\omega; t')=\int_0^\infty dte^{i\omega t}\int dq q^n\int dq' \rho(q,t+t'|q',t') q'\rho_{\rm st}(q',t').
\end{equation}
From Eq.~(\ref{eq:define_Phi_F}), the power spectrum is $\Phi(\omega)=(2/\tau_F){\rm Re}~\int_0^{\tau_F}dt' M_1(\omega;t')$.

The moments $M_n$  satisfy a set of simple linear algebraic equations
\begin{eqnarray}
\label{eq:moments_equation}
&&-i\omega M_n(\omega)+n\hat{\cal F}[M_n(\omega)]= Dn(n-1)M_{n-2}(\omega) \nonumber\\
&&+\frac{1}{2}F\left[e^{i\omega_F t'}nM_{n-1}(\omega+\omega_F) + e^{-i\omega_F t'}nM_{n-1}(\omega-\omega_F)\right] +Q_{n+1}(t').\end{eqnarray}
Here, we skipped the argument $t'$ in $M_n$ and introduced function
$
\hat{\cal F}[M_n]\equiv \kappa M_n +\beta M_{n+1} + \gamma M_{n+2}.
$. 
Functions
\begin{equation}
\label{eq:Q_n}
Q_n(t)=\int dq q^n\rho_{\rm st}(q,t)
\end{equation}  
in the right-hand side of Eq.~(\ref{eq:moments_equation}) can themselves be found from a set of linear equations similar to (\ref{eq:moments_equation}). They follow from Eq.~(\ref{eq:FP_equation}), if one sets $\rho = \rho_{\rm st}(q,t)$ and takes into account that $\rho_{\rm st}(q,t)$ is periodic in $t$. To the lowest order in $F$ it suffices to keep in $Q_n(t)$ only terms that are independent of $t$ or oscillate as $\exp(\pm i\omega_F t)$; respectively, in Eq.~(\ref{eq:Q_n}) $Q_n(t)\approx Q_n^{(0)} + \left[Q_n^{(1)}\exp(i\omega_F t)+ {\rm c.c.}\right]$, and
\begin{eqnarray}
\label{eq:Q_n_equations}
\hat{\cal F}[Q_n^{(0)}]=D(n-1)Q_{n-2}^{(0)}+F{\rm Re}\,Q_{n-1}^{(1)},\nonumber\\
i\omega_F Q_n^{(1)} + n\hat{\cal F}[Q_n^{(1)}] = Dn(n-1)Q_{n-2}^{(1)} + \frac{1}{2}nFQ_{n-1}^{(0)}.
\end{eqnarray}

The system of coupled linear equations for the moments $M_n$ and $Q_n$ can be quickly solved with conventional software to a high order in the noise intensity $D$. Nontrivial results emerge already if we keep  terms $\propto DF^2$: these are the terms that contribute to the power spectrum $\Phi_F(\omega)$ to the lowest order in $D$. To find them it suffices to consider terms $M_n$ with $n\leq 3$ and $Q_n$ with $n\leq 4$. This gives
\begin{eqnarray}
\label{eq:Phi_F_overdamped}
\Phi_F(\omega)&\approx& \frac{2D}{(\kappa^2+\omega_F^2)(\kappa^2+\omega^2)^2}\left\{2\beta^2\frac{(4\kappa^2 + \omega_F^2)(\kappa^2+ \omega^2+\omega_F^2)}{[\kappa^2+(\omega - \omega_F)^2][\kappa^2+(\omega + \omega_F)^2]}\right.
\nonumber\\
&&\left. -3\gamma\kappa\right\}. 
\end{eqnarray}
This expression refers to $|\omega|>0$; function $\Phi_F(\omega)$ contains also a $\delta$-peak at $\omega=0$, which comes from the driving-induced shift of the average static value of the coordinate. 

The solution of the equations for the moments in the considered approximation gives a correction $\propto D^2$ to the power spectrum in the absence of driving $\Phi_0(\omega)$. To the lowest order in $D$ this function displays a Lorentzian peak at $\omega=0$, $\Phi_0(\omega)=2D/(\kappa^2+\omega^2)$. This peak is used in the analysis of optical traps for Brownian particles \cite{Berg-Sorensen2004,Mas2013}; with account taken of the term $\propto D^2$ the zero-frequency peak of $\Phi_0(\omega)$ becomes non-Lorentzian.

\subsection{Power spectrum for comparatively large driving frequency}
\label{subsec:overdamped_spectrum}

The interpretation of Eq.~(\ref{eq:Phi_F_overdamped}) is simplified in the case where the driving frequency exceeds the decay rate, $\omega_F\gg \kappa$. In this case, periodic driving leads to two well-resolved features in the spectrum $\Phi_F$. One is located at $\omega=0$ and has the form 
\begin{equation}
\label{eq:small_frequency}
\Phi_F(\omega)\approx (2D/\omega_F^2)(2\beta^2-3\gamma\kappa)(\kappa^2+\omega^2)^{-2} \quad (\omega\ll \omega_F).
\end{equation}
This equation can be easily obtained directly by solving the equation of motion $\dot q=-U'(q) +F\cos\omega_Ft + f(t)$ by perturbation theory in which $q(t)$ is separated into a part oscillating at high frequency $\omega_F$ (and its overtones) and a slowly varying part. To the lowest order in $F$ and $D$, the fast oscillating part renormalizes the decay rate of the slowly varying part of $q(t)$, with $\kappa \to \kappa -(F/\omega_F)^2\left[\kappa^{-1}\beta^2 - (3/2)\gamma)\right]$. Using this correction in the expression for the power spectrum of a linear system $\Phi_0^{(0)}(\omega)=2D/(\kappa^2+\omega^2)$, one immediately obtains Eq.~(\ref{eq:small_frequency}) to the leading order in $\kappa/\omega_F$.

Interestingly, Eq.~(\ref{eq:small_frequency}) describes a peak or a dip depending on the sign of $2\beta^2-3\gamma\kappa$. That is, the sign of $\Phi_F$ is determined by the competition of the cubic and quartic nonlinearity of the potential $U(q)$. This shows high sensitivity of the spectrum to the system parameters. The typical width of the peak/dip of $\Phi_F$ near $\omega=0$ is $\kappa$; the shape of the peak/dip is non-Lorentzian. 

\begin{figure}[ht]
\center
\includegraphics[width=6cm]{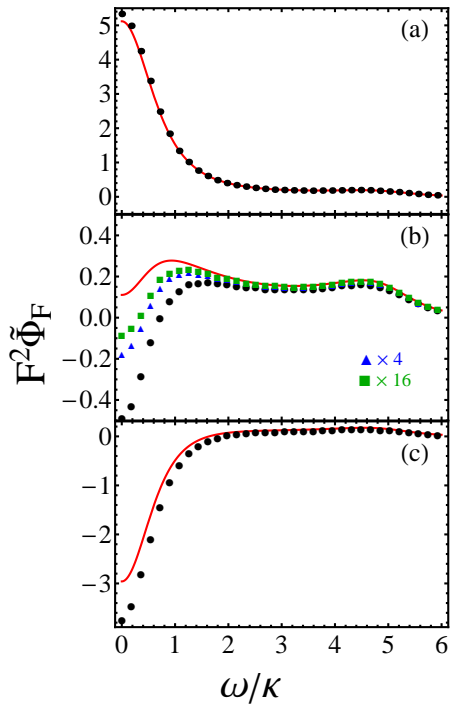}
\caption{Scaled driving induced terms in the power spectrum of an overdamped Brownian particle moving in the quartic potential $U(q)$ given by Eq.~(\ref{eq:nonlinear_potential}), $\tilde\Phi_F(\omega)=10^2\kappa^2\Phi_F(\omega)/2D$.
Panels (a), (b), and (c) refer to the scaled cubic nonlinearity $\beta^2D/\kappa^3 = 0.002$ and quartic nonlinearity $\gamma D/\kappa^2$ = 0.0006, 0.00147, and 0.002, respectively. The black dots and red solid curves correspond to the numerical simulations and Eq.~(\ref{eq:Phi_F_overdamped}). The scaled driving frequency is $\omega_F/\kappa = 5$ and the driving strength is $\kappa F^2/\omega_F^2D = 20$. For this driving strength and the noise intensity, the simulation results in panels (b) and (c) deviate from the theoretical curve. The deviation decreases for weaker driving. This is seen from the simulation data in panel (b) that refer to $\kappa F^2/\omega_F^2D = 5$ (blue triangles) and 1.25 (green squares). The corresponding  spectra are scaled up by factors 4 and 16, respectively.}. 
\label{fig:overdamped}
\end{figure}

The other spectral feature is located at $\omega_F$ and near the maximum has the form of a Lorentzian peak,  $\Phi_F(\omega)\approx (D\beta^2/\omega_F^4)[\kappa^2+(\omega - \omega_F)^2]^{-1}$. The height of this peak is  smaller by a factor $\kappa^2/\omega_F^2\ll 1$ than the height of the feature near $\omega=0$. We note that the  height of the peak at $\omega_F$ is proportional to the squared parameter of the cubic nonlinearity of the potential $\beta$, but is independent of the quartic-nonlinearity parameter $\gamma$, to the leading order in the noise intensity $D$.  

In Fig.~\ref{fig:overdamped} we compare the analytical expression (\ref{eq:Phi_F_overdamped}) with the results of numerical simulations. The simulations were done by integrating the stochastic differential equation $\dot q = -U'(q) + f(t) + F\cos\omega_Ft$ using the Heun scheme (cf. \cite{Mannella2002a}). Panel (a) shows that the cubic nonlinearity of the potential leads to a peak at $\omega=0$ and a comparatively small peak at $\omega_F$. The spectrum becomes more interesting in the generic case where both cubic and quartic terms in the potential are present and $\beta^2$ is comparable to $\gamma\kappa$. Here, as seen from panel (b), as a result of the competition between these terms, $\Phi_F(\omega)$ can have a dip at $\omega=0$ and two peaks, one near $\omega_F$ and the other with the position determined by $\beta^2/\gamma\kappa$ and $\omega_F/\kappa$. Where the quartic nonlinearity dominates, $\gamma\kappa\gg \beta^2$, see panel (c), it is hard to detect the peak at $\omega_F$ for small noise intensity. Our analytical calculations and numerical simulations show that, for larger noise intensity, this peak becomes more pronounced. 

A significant deviation of simulations and the asymptotic expression (\ref{eq:Phi_F_overdamped}) in panel (b) for small $\omega$ is a consequence of the near compensation of the contributions to $\Phi_F(\omega)$ from the cubic and quartic nonlinear terms in $U(q)$ to the lowest order in $F^2$ and $D$. The terms of higher-order in $D$ and $F^2$ become then substantial. Panel (b) illustrates how the difference is reduced if $F^2$ is reduced. We checked that by reducing also the noise intensity we obtain a quantitative agreement of simulations with Eq.~(\ref{eq:Phi_F_overdamped}).

In some cases the confining potential of an overdamped system has inversion symmetry, and then $\beta=0$ in Eq.~(\ref{eq:nonlinear_potential}). In such cases spectral features of $\Phi_F$ at the driving frequency are $\propto (\gamma D)^2$. They can be found by solving the equations for the moments $M_n$ with $n\leq 5$ and $Q_n$ with $n\leq 6$ or by solving the equations of motion by perturbation theory to the second order in $\gamma$, cf. \ref{App_susceptibility}.

\section{Power spectrum of a driven two-state system}
\label{sec:two_state}

We now consider the effect of driving on a two-state system. Various types of such systems are studied in physics, from spin-1/2 systems to two-level systems in disordered solids to classical Brownian particles mostly localized at the minima of double-well potentials. We will assume that the system dynamics are characterized by the rates $W_{ij}$ of interstate $i\to j$ switching, where $i,j=1,2$. In the case of quantum systems, this means that the decoherence rate largely exceeds $W_{ij}$; in other words, the typical duration of an interstate transition is small compared to $1/W_{ij}$. For classical systems, this description means that small fluctuations about the stable states are disregarded. 

\subsection{The model: modulated switching rates}

A major effect of periodic driving is modulation of the switching rates. It can be quite strong already for comparatively weak driving. Indeed, if the rates are determined by the interstate tunneling, since the field changes the tunneling barrier, its effect can be exponentially strong. Similarly, it may be exponentially strong in the classical limit if the switching is due to thermally activated overbarrier transitions, because the driving changes the barriers heights. Nevertheless, for weak sinusoidal driving $F\cos\omega_Ft$ the modulated rates $W_{ij}^{(F)}(t)$ can still be expanded in the driving amplitude,
\begin{equation}
\label{eq:rates_with_driving}
W_{ij}^{(F)}\approx W_{ij} -\alpha_{ij}F\cos\omega_Ft, \qquad i,j=1,2.
\end{equation}
This equation is written in the adiabatic limit, where the driving frequency $\omega_F$ is small compared to the reciprocal characteristic dynamical times, like the imaginary time of motion under the barrier in the case of tunneling \cite{Leggett1995} or the periods and relaxation times of vibrations about the potential minima in the case of activated transitions. The rates $W_{ij}$ are also assumed to be small compared to the reciprocal dynamical times. The driving frequency $\omega_F$ is of the order of $W_{ij}$.

Parameters $\alpha_{ij}$ in Eq.~(\ref{eq:rates_with_driving}) describe the response of the switching rates to the driving. They contain factors $\sim W_{ij}$. Indeed, for activated processes $W_{ij}\propto \exp(-\Delta U_i/k_BT)$, where $\Delta U_i$ is the height of the potential barrier for switching from the state $i$. If $F\cos\omega_Ft$ is the force that drives the system, then $\alpha_{ij}\approx W_{ij}d_i/k_BT$, where $d_i$ is the position  of the $i$th potential well counted off from the position of the barrier top \cite{Debye1929}.  The terms $\propto F^2$, which have been disregarded in Eq.~(\ref{eq:rates_with_driving}), are $\propto W_{ij}(d_i/k_BT)^2$ in this case; a part of these terms that are $\propto  \cos2\omega_Ft$ do not contribute to $\Phi_F(\omega)$ to the second order in $F$, whereas the contribution of the time-independent terms $\propto F^2$ comes to renormalization of the parameters $W_{ij}$ in $\Phi_0(\omega)$, see below. For incoherent interstate quantum tunneling, $\alpha_{ij}\propto W_{ij}$, too.

We will use quantum notations $|i\rangle$ ($i=1,2$) for the states of the system. One can associate these states with the states of a spin-1/2 particle by setting $|1\rangle \equiv \up$ and $|2\rangle \equiv \down$.  The system dynamics is most conveniently described by the dynamical variable $q$ defined as 
\begin{equation}
\label{eq:defined_variable}
q=|1\rangle\langle 1| -  |2\rangle\langle 2|\equiv \sigma_z,
\end{equation}
where $\sigma_z$ is the Pauli matrix. For a particle in a double-well potential, $q$ is the coordinate that takes on discrete values $1$ and $-1$ at the potential minima 1 and 2, respectively. 

The power spectra of driven two-state systems have been attracting much interest in the context of stochastic resonance,  see \cite{Dykman1995d,Wiesenfeld1998,Gammaitoni1998,Thorwart1998} for reviews. By now it has been generally accepted that, for weak driving, the power spectrum of a system has a $\delta$-peak at the driving frequency with area $\propto F^2$, which is described by the standard linear response theory \cite{Dykman1990e}. This peak is of central interest for signal processing. However, as we show in this Section, along with this peak, the spectrum has a characteristic extra structure, which is also $\propto F^2$, to the leading order in $F$.

\subsection{Kinetic equation and its general solution}

It is convenient to write the analog of Eq.~(\ref{eq:corr}) for the correlation function of the discrete variable $q$ as
\begin{equation}
\label{eq:two_state_correlator}
\langle q(t_1)q(t_2)\rangle = \sum_{i,j}\langle i|\sigma_z\hat\rho(t_1|t_2)\sigma_z\hat\rho_{\rm st}(t_2)|j\rangle 
\end{equation}
Here, $\hat\rho(t_1|t_2)$ is the transition density matrix, $\hat\rho(t_1|t_2)\equiv \sum|i\rangle \rho_{ij}(t_1|t_2)\langle j|$, and $\hat\rho_{\rm st}\equiv \sum |i\rangle (\rho_{\rm st})_{ii}  \langle i|$ is the stationary density matrix. By construction (in particular, because of the decoherence in the quantum case) the stationary density matrix is diagonal. Its matrix elements $(\rho_{\rm st})_{ii}$ give the populations of the corresponding states and periodically depend on time,   $\hat\rho_{\rm st}(t+2\pi/\omega_F)=\hat \rho_{\rm st}(t)$. The transition matrix elements $ \rho_{ij}(t_1|t_2)$ give the probability to be in state $i$ at time $t_1$ given that the system was in state $j$ at time $t_2$. At equal times we have  $\hat\rho(t_2|t_2) = \hat I$, where $\hat I$ is the unit matrix. 

Equation~(\ref{eq:two_state_correlator}) does not have the form of a trace over the states $|i\rangle$; rather it expresses the correlator in terms of the joint probability density to be in state $|j\rangle$ at time $t_2$ and in state $|i\rangle$ at time $t_1$, with summation over $i,j$ \cite{vanKampen_book}. In the quantum formulation, the applicability of this expression is a consequence of the decoherence and Markovian kinetics.

Matrix elements $\rho_{ij}(t|t')$  satisfy a simple balance equation, which in the presence of driving reads
\begin{equation}
\label{eq:balance_two_state}
\partial_{t}\rho_{1j}(t|t')= -W_{12}^{(F)}(t)\rho_{1j} + W_{21}^{(F)}(t)\rho_{2j},\quad \rho_{1j} + \rho_{2j} = 1,
\end{equation}
where $j=1,2$. Equation for the matrix elements of $\hat\rho_{\rm st}(t)$ has the same form, except that subscript $j$ has to be set equal to the first subscript.

From Eqs.~(\ref{eq:two_state_correlator}) and (\ref{eq:balance_two_state}) we obtain a general expression for the correlator of interest,
\begin{eqnarray}
\label{eq:two_state_general}
&&\langle q(t_1)q(t_2)\rangle =  \exp\left[-\int_{t_2}^{t_1}dt W^{(F)}_+(t)\right]+  \langle\sigma_z(t_2)\rangle_{\rm st}  \int_{t_2}^{t_1}dt \left\{W^{(F)}_-(t)
\nonumber \right.\\
 &&\left. \times\exp\left[-\int_{t}^{t_1}dt' W^{(F)}_+(t')\right]\right\}, \qquad 
W^{(F)}_\pm (t) = W_{21}^{(F)}(t) \pm W_{12}^{(F)}(t).
\end{eqnarray}
Here, $\langle \sigma_z(t)\rangle_{\rm st} \equiv \langle q(t)\rangle_{\rm st} \equiv {\rm Tr}~[\sigma_z \hat\rho_{\rm st}(t)]$ is the time-dependent difference of the state populations in the stationary state. 
Generally , $\langle \sigma_z(t)\rangle_{\rm st}$ is nonzero even in the absence of driving unless the switching rates are equal, $W_{12}=W_{21}$. In the presence of driving there emerges a periodic term in $\langle \sigma_z(t)\rangle_{\rm st}$, which describes the linear response, for weak driving.  

Disregarding terms oscillating as $\exp(\pm 2 i\omega_F t)$, to the second order in $F$ we obtain from the balance equation (\ref{eq:balance_two_state}) written for $(\rho_{\rm st})_{ii}$
\begin{eqnarray}
\label{eq:stationary_sigma}
&&\langle \sigma_z(t)\rangle_{\rm st} \approx \frac{W_-}{W_+} + \frac{F}{2}\left [\chi_1(\omega_F)e^{-i\omega_F t} + {\rm   c.c.}\right] + \frac{\alpha_+ F^2}{2W_+} {\rm Re}\, \chi_1(\omega_F), \nonumber \\
&& \chi_1(\omega) = 2\left(\alpha_{12} W_{21} - \alpha_{21} W_{12}\right)/\left[W_+\left(W_+ - i\omega\right)\right].
\end{eqnarray}
Here we introduced notations
\begin{equation}
\label{eq:pm_notations}
\alpha_\pm = \alpha_{21} \pm \alpha_{12}, \qquad W_\pm = W_{21}\pm W_{12}.
\end{equation}
Function $\chi_1(\omega)$ gives the linear susceptibility. In the case of thermally activated transitions, Eq.~(\ref{eq:stationary_sigma}) for $\chi_1$ coincides with the classical result \cite{Debye1929}. The term $W_-/W_+$ gives the difference of the state populations in the absence of driving, whereas the term $\propto F^2$ gives the time independent part of the driving-induced correction to this difference.

\subsection{The driving-induced part of the power spectrum}
\label{subsec:power_spectrum_two_state}

Equation~(\ref{eq:two_state_general}) allows one to calculate the period-averaged correlator $\lal q(t_1)q(t_2)\rar$ in the explicit form and to obtain the power spectrum. As before, we will not consider the $\delta$-peak in $\Phi(\omega)$ for $\omega=0$. The spectrum is an even function of $\omega$, and we will consider it for $\omega > 0$:
\begin{eqnarray}
\label{eq:two_state_power}
&\Phi_0(\omega)= 8\frac{W_{12}W_{21}}{W_+^2}\,\frac{W_+}{W_+^2 + \omega^2}, \qquad \Phi_F(\omega) = \Phi_F^{\rm(r)}(\omega) + \Phi_F^{\rm (c)}(\omega), \nonumber\\
&\Phi_F^{\rm(r)}(\omega)=\alpha_+\sum_{\mu,\nu=\pm}\phi_F(\mu\omega,\nu\omega_F),\nonumber\\
&\phi_F(\omega,\omega_F) = - [W_+   -  i(\omega-\omega_F)]^{-1}\left[\frac{\alpha_+ W_{12}W_{21}}{\omega_F^2W_+^2} + i\frac{W_-}{2\omega_FW_+}\chi_1^*(\omega_F)\right].
\end{eqnarray}
The term $\Phi_0$ is the familiar power spectrum of a two-state system in the absence of driving \cite{Debye1929}. It has a peak at $\omega=0$ with halfwidth $W_+$ equal to the sum of the switching rates. The term $\Phi_F^{\rm (c)}$ describes the driving-induced modification of the peak centered at $\omega=0$, 
\begin{eqnarray}
\label{eq:zero_frequency_peak_two_state}
\Phi_F^{\rm (c)}(\omega) = (\alpha_+^2/2\omega_F^2)\Phi_0(\omega) -
|\chi_1(\omega_F)|^2W_+/(W_+^2+\omega^2) .
\end{eqnarray}

Of major interest to us is the part $\Phi_F^{\rm (r)}(\omega)$ of the driving-induced term in the power spectrum (\ref{eq:two_state_power}). For $\omega>0$, it shows a resonant peak (or a dip, depending on the parameters) at the driving frequency $\omega_F$. In contrast to the $\delta$-peak of the linear response, the peak has a finite halfwidth $\sim W_+= W_{12}+W_{21}$. It is well separated from the peak at $\omega=0$ for $\omega_F\gg W_+$ and generally is of a non-Lorentzian shape. We stress that, to the order of magnitude, the peak has the same overall area as the $\delta$-peak of the linear response (in the case of a dip, the absolute value of the area should be considered). Another important feature of the peak/dip seen from Eq.~(\ref{eq:two_state_power}) is that it is proportional to the parameter $\alpha_+=\alpha_{12}+\alpha_{21}$. This parameter describes the change of the sum of the switching rates due to the driving.  

For activated switching between potential minima considered in the classical stochastic resonance theory, $\alpha_+ = (k_BT)^{-1}(W_{12}d_1+ W_{21} d_2)$. For a symmetric potential $\alpha_+=0$, since $W_{12}=W_{21}$ and $d_1 = -d_2$.  Then $\Phi_F^{\rm (r)}=0$, in agreement with \cite{Mcnamara1989} where a symmetric potential was considered. On the other hand, for strong driving it was found \cite{Nikitin2007} that the power spectrum for an asymmetric potential displays peaks close to odd multiples of the driving frequency and dips close to even multiples of driving frequency. In our weak-driving analysis we do not consider peaks/dips near the overtones of $\omega_F$; however, as seen from Eq.~(\ref{eq:two_state_power}), the sign of $\Phi_F^{\rm (r)}(\omega)$ near $\omega_F$ can be positive or negative, depending on the parameters.

\begin{figure}[h]
\center
\includegraphics[width=11cm]{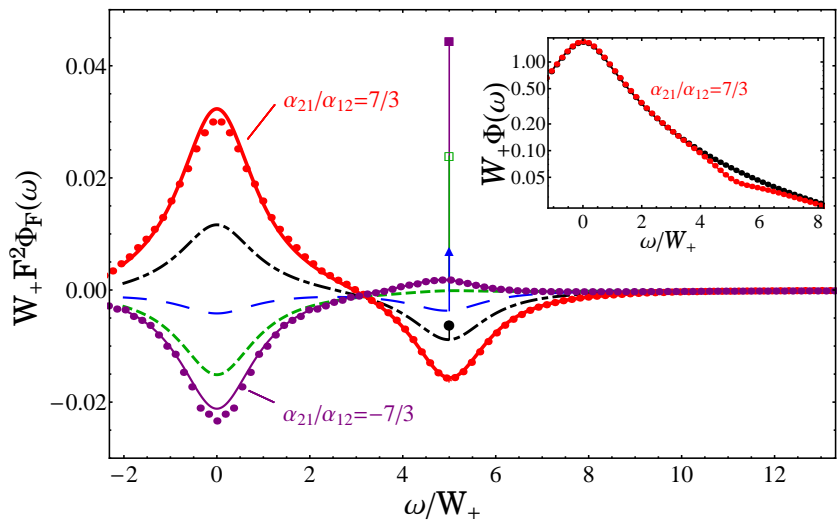} 
\caption{The driving induced terms in the power spectrum of the two-state system for the ratio of the switching rates  $W_{21}/W_{12} = 7/3$. The scaled driving frequency and amplitude are  $\omega_F/W_+ = 5$ and $F\alpha_{12}/W_{12} = 1.$ On the thick solid (red), dot-dashed (black), long-dashed (blue), short-dashed (green), and thin solid (purple) lines the ratio  $\alpha_{21}/\alpha_{12}$ is 7/3, 7/6, 0 , $-7/6$,  and $-7/3$. The vertical line at $\omega_F$ shows the position of the $\delta$-peak at $\omega_F$. The areas of the  $\delta$-peaks for different $\alpha_{21}/\alpha_{12}$ are given by the heights of the vertical segments. The heights are counted off from the lines to the symbols of the same color, i.e., to the circle, triangle, and open and full square,  in the order of decreasing $\alpha_{21}/\alpha_{12}$; there is no symbol for  $\alpha_{21}/\alpha_{12}=7/3$ as there is no $\delta$-peak in this case. The inset shows the full spectrum with (red) and without (black) driving for $\alpha_{21}/\alpha_{12} = 7/3$. The curves and the dots show the analytical theory and the simulations, respectively.}
\label{fig:ts}
\end{figure}

Examples of the driving-induced spectra $\Phi_F(\omega)$ are shown in Fig.~\ref{fig:ts}. One can clearly see the peaks or dips both at $\omega=0$ and at the driving frequency $\omega_F$.  In agreement with Eqs.~(\ref{eq:two_state_power}) and (\ref{eq:zero_frequency_peak_two_state}), the signs of the features of $\Phi_F$ are determined by  the interrelation between the parameters of the two-states system. For illustration purpose we chose the values of the ratio of the response parameters $\alpha_{21}/\alpha_{12}$ to lie between plus and minus the ratio of the switching rates in the absence of driving, $W_{21}/W_{12}$. As seen from Fig.~\ref{fig:ts}, the spectra are very sensitive  to the ratio $\alpha_{21}/\alpha_{12}$. We have seen this sensitivity also for different values of $W_{21}/W_{12}$. 

Unexpectedly, a finite-height spectrum $\Phi_F(\omega)$ emerges even where the linear susceptibility is equal to zero, which happens for $\alpha_{12}W_{21}=\alpha_{21}W_{12}$. This is seen from Eq.~(\ref{eq:two_state_power}) and also from Fig.~\ref{fig:ts}. The red line with $\alpha_{21}/\alpha_{12}=7/3$ refers to this case, and the area of $\delta$-peak in the spectrum is zero. As seen from the figure, numerical simulations are in excellent agreement with the analytical expressions.

The structure of the spectrum near $\omega=0$ will be modified if one takes into account terms $\propto F^2$ in the expressions for the switching rates (\ref{eq:rates_with_driving}). In the considered leading-order  approximation in $F$ these terms have to be averaged over the driving period and are thus independent of time. The correction due to these terms can be immediately found from Eq.~(\ref{eq:two_state_power}) for $\Phi_0(\omega)$ by expanding $\Phi_0$ to the first order in the corresponding increments of  $W_{ij}$; this correction is of a non-Lorentzian form.

\section{Threshold detector}
\label{sec:threshold}

An insight into the dynamical nature of the driving-induced change of the power spectrum can be gained from the analysis of the spectrum of a threshold detector. Such detectors are broadly used in science and engineering, and their analogs play an important role in biosystems. We will employ the simplest model where the output of a threshold detector is $q=-1$ if the signal at the input is below a threshold value $\eta$, whereas $q=1$ otherwise, and will consider the case where the input signal is a sum of the periodic signal $F\cos\omega_Ft$ and noise $\xi(t)$, 
\begin{eqnarray}
\label{eq:threshold_model}
q(t) = 2\Theta\left[F(t)+\xi(t) - \eta\right] - 1,
\end{eqnarray}
where $\Theta(x)$ is the Heaviside step function. To avoid singularities related to non-differentiability of the $\Theta$-function, we will model the output by 
\begin{equation}
\label{eq:tanh_model}
q(t)=\tanh\left[\Lambda\bigl(F(t)+\xi(t)-\eta\bigr)\right], \qquad \Lambda \gg 1,
\end{equation}
and in the final expressions will go to the limit $\Lambda\to\infty$. Much work on the interplay of noise and driving in threshold detectors has been done in the context of stochastic resonance, cf. \cite{Wiesenfeld1994,Gingl1995a,Jung1995}. In these papers of primary interest was the signal to noise ratio; the issues we are considering here, i.e., the occurrence of the effective ``inelastic scattering" and ``fluorescence" as a result of interplay of nonlinearity and noise, have not been addressed, to the best of our knowledge.

In the absence of noise, the power spectrum of $q(t)$ is a series of $\delta$-peaks at $\omega_F$ and its overtones (including $\omega=0$), provided the driving amplitude $F>\eta$, whereas for $F<\eta$ we have $q=-1$ and the power spectrum is just a $\delta$-peak at $\omega=0$. On the other hand, if $F=0$ and $\xi(t)$ is white noise, in the limit $\Lambda\to\infty$ in Eq.~(\ref{eq:tanh_model}) the correlator $\langle q(t)q(t')\rangle =0$ for $t\neq t'$, since the values of $q(t)$ at different instants of time are uncorrelated and $\langle q\rangle \to 0$, whereas $\langle q^2\rangle \to 1$.

The singular behavior of the correlator $\langle q(t)q(t')\rangle$ in the case of white noise persists also in the presence of driving. This is a consequence of the absence of {\it dynamics}, i.e., any memory effects in the variable $q(t)$ (\ref{eq:threshold_model}), and the singular distribution of white noise, where the intensity $\langle \xi^2(t)\rangle$ diverges.  

Dynamics can be brought into the system by the noise color. Such noise can be thought of as coming from a dynamical system with retarded response, which is driven by white noise.  We will be interested in the correlator $\langle q(t)q(t')\rangle$ and the power spectrum $\Phi(\omega)$  for weak driving, where the driving amplitude is $F\ll \eta$ (subthreshold driving), and for a simple colored noise, the Ornstein-Uhlenbeck noise. This is Gaussian noise with correlator 
\begin{equation}
\label{eq:color_noise}
\langle \xi(t)\xi(t')\rangle = (D/\kappa) \exp(-\kappa|t-t'|).
\end{equation}
Parameter $\kappa$ characterizes the decay rate of noise correlations. 

Because the threshold detector has no dynamics on its own, the value of the variable $q(t)$ is determined by the instantaneous value of the noise $\xi(t)$. We can write $q(t)\equiv \tilde q\bigl(t,\xi(t)\bigr)$, where $\tilde q(t,\xi)$ is given by Eqs.~(\ref{eq:threshold_model}), (\ref{eq:tanh_model}) with $\xi(t)$ replaced with $\xi$. Then the general expression for the correlator of $q(t)$, Eq.~(\ref{eq:corr}), can be rewritten as
\begin{eqnarray}
\label{eq:td_corr}
\la  q(t_1)q(t_2) \ra = \int d\xi_1 d\xi_2\, \tilde q(t_1, \xi_1)\tilde q(t_2,\xi_2)\,
\rho^{(\xi)}(\xi_1,t_1|\xi_2,t_2)\rho^{(\xi)}_{\rm   st}(\xi_2,t_2).
\end{eqnarray}
Here, the superscript $\xi$ indicates that the corresponding transition probability density and the stationary distribution refer to the random process $\xi(t)$. 

The form of the transition probability for the process (\ref{eq:color_noise}) is well-known \cite{Risken1989}, 
\begin{eqnarray}
\label{eq:transition_prob_xi}
\fl
\rho^{(\xi)}(\xi_1,t_1|\xi_2,t_2) = \sqrt{\frac{\kappa}{2\pi D(1-e^{-2\kappa|t_1-t_2|})}} 
\exp\left\{-\frac{\kappa(\xi_1-\xi_2 e^{-\kappa|t_1-t_2|})^2}{2D(1-e^{-2\kappa|t_1-t_2|})}\right\}.
\end{eqnarray}
The stationary distribution $\rho^{(\xi)}_{\rm st}(\xi_1,t_1)$ is given by the same expression with $t_2\to -\infty$.  Substituting these expressions into Eq.~(\ref{eq:td_corr}) and expanding $\tilde q(t,\xi)$ in $F(t)$, after averaging over the driving period  we obtain to second order in $F(t)$ for $t_1>t_2$
\begin{eqnarray}
\label{eq:threshold_correlator}
\fl
\lal  q(t_1)q(t_2) \rar  =C + & 4\int_{\eta }^{\infty} d\xi_1 \int_{\eta}^{\infty} d\xi_2\left[ \rho^{(\xi)}(\xi_1,t_1|\xi_2,t_2) -
 \rho^{(\xi)}_{\rm st}(\xi_1)\right] \rho^{(\xi)}_{\rm   st}(\xi_2) \nonumber\\
&+  2F^2 \cos\wF(t_1-t_2)  \rho^{(\xi)}(\eta,t_1|\eta,t_2)  \rho^{(\xi)}_{\rm   st}(\eta)\nonumber\\
 &- 2F^2\int_{\eta}^{\infty}d\xi_2\rho^{(\xi)}_{\rm   st}(\xi_2) \frac{d}{d\eta}\left[\rho^{(\xi)}(\eta,t_1|\xi_2,t_2) - 
\rho^{(\xi)}_{\rm   st}(\eta)\right].
\end{eqnarray}
Here, $C$ is a constant independent of time; it leads to a $\delta$ peak at $\omega = 0$ in the power spectrum and will not be considered in what follows. The remaining terms are time-dependent. They decay with increasing $|t_1-t_2|$, except for the term that oscillates as $\exp[\pm i\omega_F(t_1-t_2)]$ and describes the standard linear response to periodic driving. As seen from Eq.~(\ref{eq:threshold_correlator}), this term has the form 
\begin{eqnarray}
\label{eq:susceptibility_threshold}
&2F^2 \cos\wF(t_1-t_2)\left[\rho^{(\xi)}_{\rm   st}(\eta)\right]^2 \equiv 
\frac{1}{2}F^2|\chi(\omega_F)|^2\cos\omega_F(t_1 - t_2) , \nonumber\\
&\chi(\omega)= 2\rho^{(\xi)}_{\rm   st}(\eta)\equiv (2\kappa/\pi D)^{1/2}\exp[-\kappa\eta^2/2D),
\end{eqnarray}
where $\chi(\omega)$ is the standard linear susceptibility \cite{LL_statphys1}  of the threshold detector. Interestingly, this susceptibility is independent of frequency. This is because the detector has no dynamics, its response to the driving is instantaneous. An alternative derivation of the expression for the susceptibility, which provides a useful insight  into the response of the threshold detector, is given in \ref{App_susceptibility}. It also shows how to deal with the singularities in Eq.~(\ref{eq:threshold_correlator}) for $t_1\to t_2$, which emerge after the transition $\Lambda\to\infty$ in Eqs.~(\ref{eq:tanh_model}) and (\ref{eq:td_corr}).

The power spectrum $\Phi(\omega)$ is obtained from Eq.~(\ref{eq:threshold_correlator}) by a Fourier transform. The $F$-independent term in Eq.~(\ref{eq:threshold_correlator}) gives the power spectrum $\Phi_0(\omega)$ in the absence of driving. It has a peak at $\omega=0$. The term $\propto \cos\omega_F(t_1 - t_2)$ gives a $\delta$-peak and also a finite-width peak $F^2\Phi_F^{\rm (r)}(\omega)$ at frequency $\omega_F$. The last term in Eq.~(\ref{eq:threshold_correlator}) gives a driving-induced  feature in the power spectrum at zero frequency $F^2\Phi^{\rm (c)}(\omega)$.

The shape of the spectra is determined by the dimensionless parameter that characterizes the ratio of the threshold  to the noise amplitude $\eta(\kappa/D)^{1/2}$. For weak noise, where  $\eta(\kappa/D)^{1/2}\gg 1$,  the peak near $\omega_F$ has the form
\begin{equation}
\label{eq:threshold_asymptotic_resonant}
\Phi_F^{\rm (r)}(\omega) \approx \frac{1}{D\sqrt{2\pi}}\,{\rm Re}\,\left(\frac{\kappa\eta^2}{4D}+i\frac{\omega - \omega_F}{\kappa}\right)^{-1/2}e^{-\kappa\eta^2/2D}.
\end{equation}
Here we assumed that $\omega_F/\kappa$ is sufficiently large, so that the features of $\Phi_F$ centered at $\omega_F$ and $\omega=0$ are well separated; Eq.~(\ref{eq:threshold_asymptotic_resonant}) applies for $|\omega-\omega_F|\ll \omega_F$. The spectrum (\ref{eq:threshold_asymptotic_resonant}) has a characteristic non-Lorentzian form with typical width $\kappa^2\eta^2/4D$. However, its area is small. 

In the opposite limit of low threshold, $\eta(\kappa/D)^{1/2}\ll 1$, to the leading order 
\begin{equation}
\label{eq:threshold_resonant_small}
 \Phi_F^{\rm (r)}(\omega)\approx \frac{1}{2\sqrt{\pi}D}\,{\rm Re}\left[\Gamma\left(i\frac{\omega-\omega_F}{2\kappa}\right)/\Gamma\left(\frac{1}{2}+i\frac{\omega-\omega_F}{2\kappa}\right) \right]
\end{equation}
near $\omega_F$. This spectrum falls off slowly away from the maximum, as $|\omega-\omega_F|^{-1/2}$ for $|\omega-\omega_F|\gg \kappa$.  Equation~(\ref{eq:threshold_resonant_small}) does not contain the threshold $\eta$. The small-$\eta$ correction to  (\ref{eq:threshold_resonant_small}) for $\omega=\omega_F$ is $(1-\ln 2)\kappa\eta^2/\pi D^2$. It is positive. From the comparison of Eqs.~(\ref{eq:threshold_asymptotic_resonant}) and (\ref{eq:threshold_resonant_small}), one sees that the height of the peak at $\omega_F$ first increases with the increasing $\eta(\kappa/D)^{1/2}$, but then starts decreasing.

\begin{figure}[h]
\includegraphics[width=74mm]{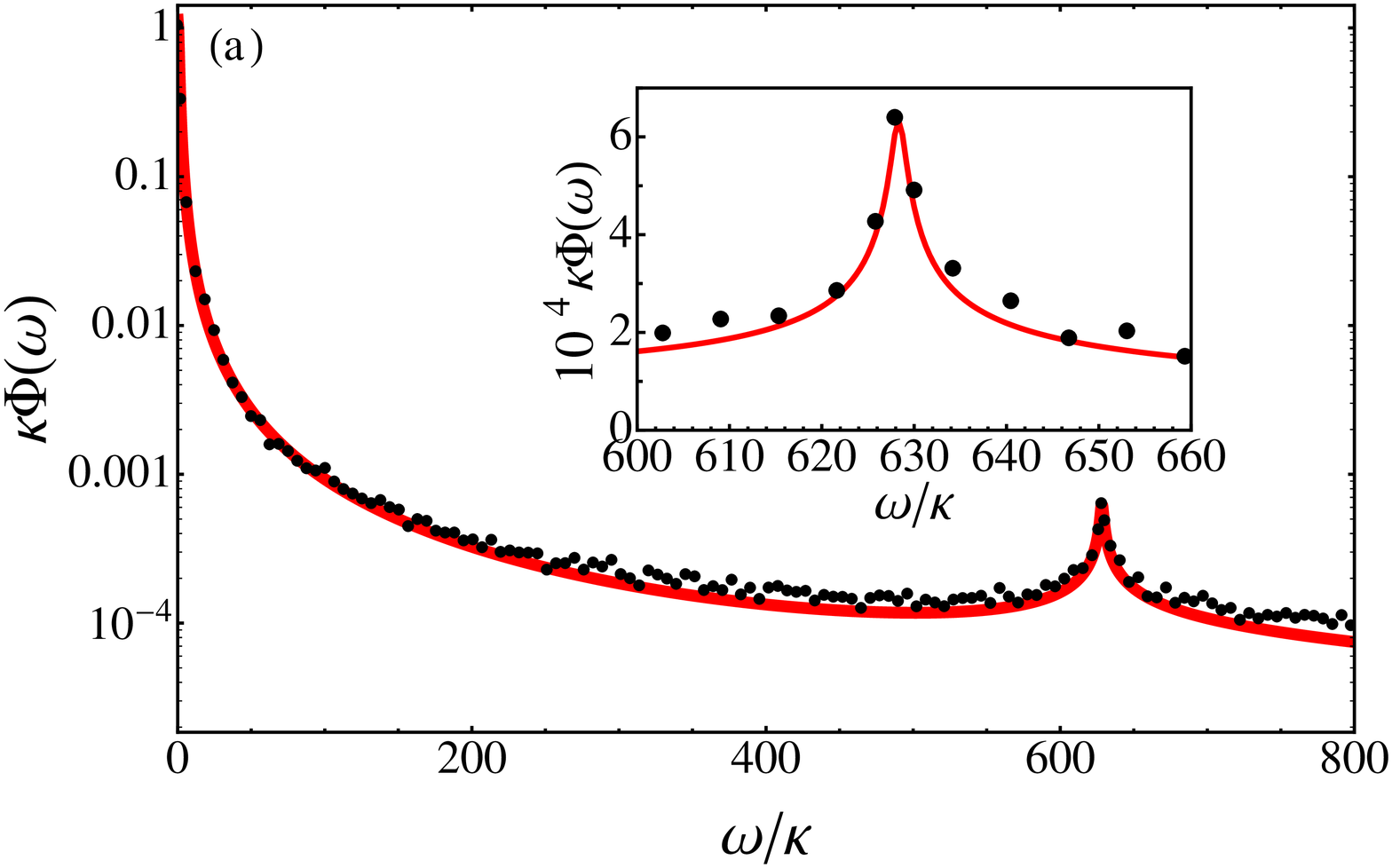}
\includegraphics[width=74mm]{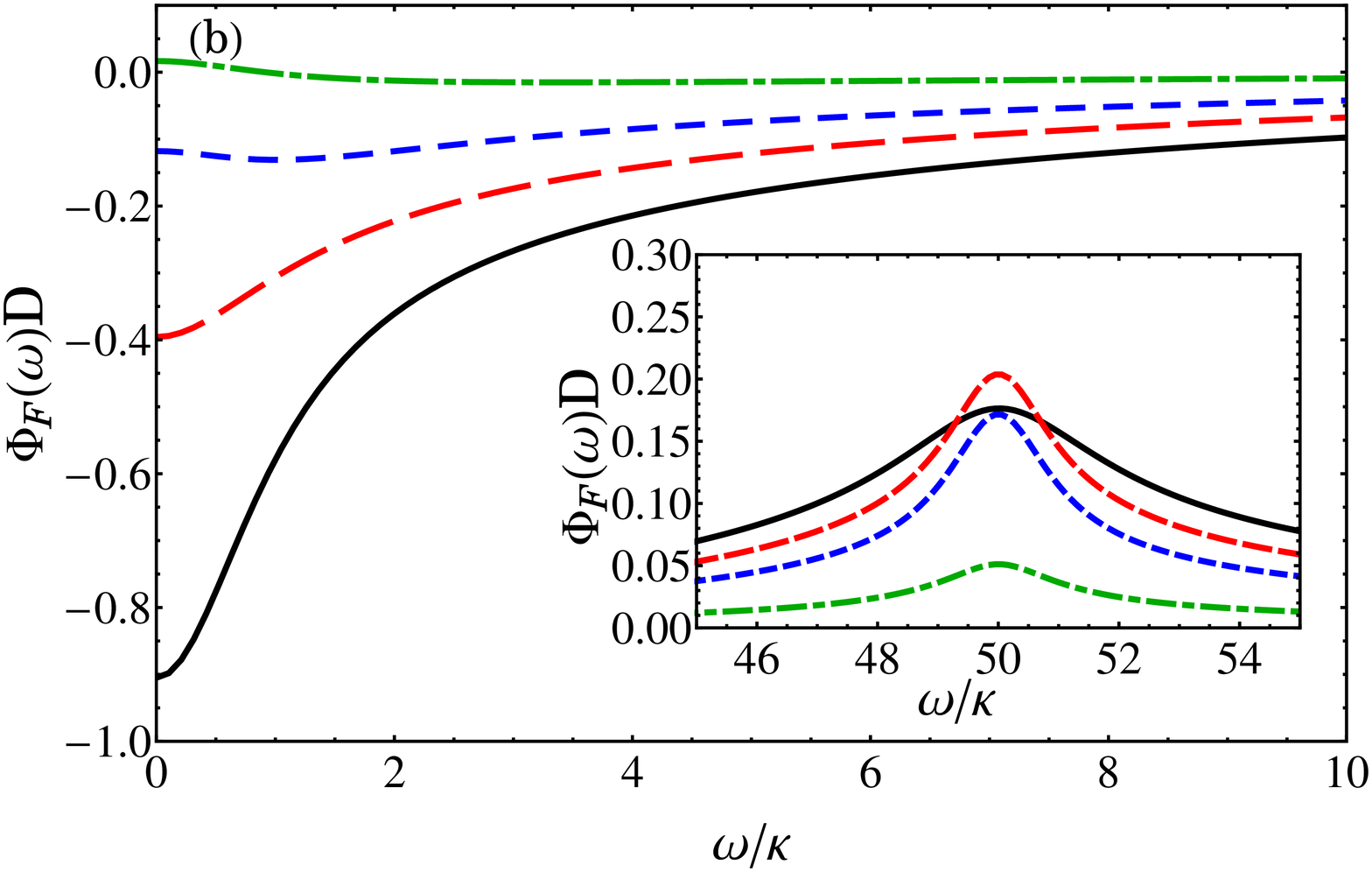}
\caption{Power spectrum of the threshold detector. (a): The full power spectrum; the scaled frequency and the intensity of the driving are $\wF /2\pi\kappa = 100$ and  $F^2\kappa/D = 0.0025$. The scaled threshold is $\eta(\kappa/D)^{1/2}  = 0.5$. Inset: the spectrum near the driving frequency. The delta peak has been subtracted. The curves and black dots refer to the theory and simulations, respectively. (b): The low-frequency part of the driving-induced term in the power spectrum for $\omega_F/\kappa = 50$ as given by Eq.~(\ref{eq:threshold_correlator}). The solid (black), long-dashed (red), short-dashed (blue) and dot-dashed (green) curves correspond to  the scaled value of the threshold $\eta( \kappa/D)^{1/2} = 0.1, 0.8, 1.2$, and 2. Inset: the spectrum near the driving frequency, $\wF /\kappa =50 $.  } 
\label{fig:td}
\end{figure}

In Fig.~\ref{fig:td} we show analytical results for the power spectra obtained from  Eq.~(\ref{eq:threshold_correlator}) for several parameter values and compare them with the results of simulations. Immediately seen from this figure is that the driving modifies the overall spectrum  most significantly near $\omega=0$ and near $\omega_F$ for large $\omega_F/\kappa$. There emerges a finite-width peak at $\omega_F$. As seen from the inset in panel (b), the width of this peak increases with decreasing noise intensity, that is, with increasing $\eta (\kappa/D)^{1/2}$. This is a counterintuitive consequence of the unusual interplay of noise and driving in a threshold detector. The height of the peak displays a nonmonotonic dependence on  $\eta (\kappa/D)^{1/2}$. 

The low-frequency spectrum $\Phi_F(\omega)\approx \Phi_F^{\rm (c)}(\omega)$ also displays a pronounced feature near $\omega=0$. One can show from the analysis of the last term in Eq.~(\ref{eq:threshold_correlator}) that, for small $\eta(\kappa/D)^{1/2}$, this feature is a dip, with $\Phi_F^{\rm (c)}(0)=-1/D$ for $\eta(\kappa/D)^{1/2}\to 0$. The shape of the dip is non-Lorentzian, with typical width $\kappa$.  As $\eta(\kappa/D)^{1/2}$ increases, the depth of the dip decreases. Ultimately the shape of the spectrum changes completely. For large $\eta(\kappa/D)^{1/2}$ the spectrum $\Phi_F^{\rm (c)}$ becomes broad and shallow. To the leading order in $[\eta(\kappa/D)^{1/2}]^{-1}$, it can be written as $(2/\pi D)(D/ \kappa\eta^2)^{1/2} \exp(-\kappa\eta^2/2D)\tilde\Phi_F^{\rm (c)}(2D\omega/\kappa^2\eta^2)$, where the dimensionless function $\tilde\Phi_F^{\rm(c)} (x)$ is zero for $x=0$, has a minimum at $x\approx 1.7$, where it is $\approx -0.6$, and then approaches zero with increasing $x$ as $x^{-1/2}$.

\section{Conclusions}

The results of this paper demonstrate that the interplay of driving and fluctuations leads to the onset of specific spectral features in the power spectra of dynamical systems. Such features are analogs of inelastic light scattering and fluorescence in optics, where an electromagnetic field can excite radiation at a frequency shifted from its frequency and also at the characteristic system frequency. Our results show that, in classical systems and in incoherent quantum systems, the spectral features emerge as a result of the fluctuation-induced modulation of the response to the driving. Such modulation is common to nonlinear systems. 

Since nonlinearity and noise are always present in real systems, the occurrence of the driving-induced spectral features in the power spectra should be also generic. However, these features are specific for particular systems, which allows using them for system characterization.

We have studied three types of systems, all of which are attracting significant interest in mesoscopic physics and in several other areas of science. The first one is an overdamped Brownian particle fluctuating in a nonparabolic potential well. This model describes, in particular, small particles and molecules optically trapped in a liquid. We find that, when the particle is periodically driven, the nonparabolicity of the potential leads to an extra spectral peak or a dip at zero frequency. For comparatively weak noise, the sign of the driving-induced term in the spectrum at small $\omega$ is determined by the competition of the cubic and quartic nonlinearity of the potential. The overall shape of the low-frequency spectrum strongly depends on the form of the confining potential as well. In addition, along with a $\delta$-peak at the driving frequency, the driving-induced spectrum displays a peak at this frequency with a width of the order of the relaxation rate of the system. 

We have also studied a two-state system that at random switches between the states. We assumed that the driving modulates the rates of interstate switching. The driving-induced spectrum has a rich form. Depending on the interrelation between the switching rates without driving and the driving-induced corrections to the rates, it can have peaks or dips both at $\omega=0$ and at the driving frequency. The typical width of the peaks/dips is given by the sum of the interstate switching rates without driving. Interestingly, these finite-width spectral features can emerge even where the $\delta$-peak at the driving frequency has very small (or zero) intensity.

The third system we studied is a threshold detector. Here the dynamical nature of the driving-induced spectral change is particularly pronounced, as this change does not occur if the noise in the detector is  white, except for the $\delta$-peak at the driving frequency. On the other hand, for colored noise driving does change the power spectrum nontrivially. As in other systems, we find a driving-induced spectral feature near zero frequency. It can be a peak or a dip depending on the ratio of the threshold to the appropriately scaled noise intensity. Also, the height of the finite-width peak at the driving frequency displays a nonmonotonic dependence on this ratio, as does the width of the peak, too, i.e., noise can both increase or decrease the width.

In all studied systems inertial effects played no role: the peaks of the power spectra are located at zero frequency in the absence of driving. Therefore driving-induced spectral features near the driving frequency and zero frequency correspond to inelastic scattering and fluorescence, respectively. However, in contrast to the conventional fluorescence, driving can induce a dip in the spectrum at zero frequency, as we have seen in all studied systems (the total power spectrum remains positive, of course). The occurrence of the dip looks as if the driving were decreasing the noise in the system, although in fact the dip has dynamical nature.

The power spectra of weakly damped nonlinear systems should also display extra features in the presence of weak periodic driving. The effect should be most pronounced where the driving is resonant. Along with the features near the driving frequency and near $\omega=0$, there should arise features near the eigenfrequencies of slowly decaying vibrations about the stable states. Several features of the power spectra have been studied for nonlinear  oscillators in the regime of strong driving, see recent papers \cite{Leyton2012,Dykman2011} and references therein. Interestingly, the results do not immediately extend to the weak-driving regime, and the features of the interplay of nonlinearity and driving where they are of comparable strength remain to be explored. However, it is clear from the presented results that the driving-induced change of the spectra is a general effect that provides a sensitive tool for characterizing fluctuating systems and their parameters. 

The research of YZ and MID was supported in part by the US Army Research Office (grant W911NF-12-1-0235), US Defense Advanced Research Agency (grant FA8650-13-1-7301), and by TOYOTA Central R\&D Labs., Inc.


\appendix
\section{Formulation in terms of fluctuating susceptibilities}
\label{App_susceptibility}

The change of the power spectrum induced by the driving can be analyzed in terms of the fluctuating linear and nonlinear susceptibility of the system. If the dynamical variable that describes the state of the system is $q(t)$, to the second order in the driving $F(t)$ we have
\begin{eqnarray}
\label{eq:susceptibility}
\fl
q(t) \approx q_0(t)+\int_{-\infty}^tdt' \chi_1(t,t')F(t') + \int\!\!\!\int_{-\infty}^t dt' dt''\chi_2(t,t',t'')F(t')F(t''),
\end{eqnarray}
where $q_0(t)$ is the (random) value of $q(t)$ in the absence of driving. Functions $\chi_1$ and $\chi_2$ describe the linear and nonlinear response. We emphasize that these functions themselves are random, there is no ensemble averaging in Eq.~(\ref{eq:susceptibility}). This equation is merely a consequence of the causality principle. Spatial and temporal fluctuations of the linear susceptibility $\chi_1$ are standardly considered in the context of light scattering \cite{Einstein1910,Reichl2009}. However, the analysis of the power spectrum of nonlinear systems in the presence of driving requires also taking into account the fluctuating nonlinear susceptibility $\chi_2$.

The linear and nonlinear fluctuating susceptibilities lead to two terms in the driving-induced part of the power spectrum defined by Eq.~(\ref{eq:define_Phi_F}),  $\Phi_F(\omega)= \Phi_F^{(1)}(\omega) + \Phi_F^{(2)}(\omega)$. Substituting into (\ref{eq:susceptibility}) $F(t) = F\cos\omega_Ft \exp(\epsilon t)$ with $\epsilon\to +0$, we obtain  \cite{Zhang2014}
\begin{eqnarray}
\label{eq:Phi_1_general}
\Phi_F^{(1)}(\omega)&=& \frac{1}{2}{\rm Re}\int_0^\infty dt e^{i(\omega-\omega_F)t}
\int\!\!\!\int_{-\infty}^0 dt' dt'' e^{i\omega_F(t'' - t')}\nonumber\\
&&\times
\Big\langle\chi_1(t,t+ t')[\chi_1(0,t'')-\langle \chi_1(0,t'')\rangle]\Big\rangle,
\end{eqnarray}
and
\begin{eqnarray}
\label{eq:Phi_2_general}
\Phi_F^{(2)}(\omega)&=& {\rm Re}\int_0^\infty dt e^{i\omega t}\int\!\!\!\int_{-\infty}^0 dt' dt'' \cos[\omega_F(t'-t'')]
\nonumber\\
&&\times \left[\langle \chi_2(t,t+t',t+t'')q_0(0)\rangle + \langle q_0(t)\chi_2(0,t',t'')\rangle\right],
\end{eqnarray}
where we assumed $\langle q_0(t)\rangle = 0$. 

The general form of Eqs.~(\ref{eq:Phi_1_general}) and (\ref{eq:Phi_2_general})  immediately shows two distinct effects of the driving, which are pronounced where the driving frequency $\omega_F$ largely exceeds the typical relaxation rate of the system. The term $\Phi_F^{(1)}(\omega)$ is a function of the detuning of frequency $\omega$ from the driving frequency  $\omega_F$. Therefore one may expect that $\Phi_F^{(1)}(\omega)$ displays features like peaks or dips near $\omega_F$. In contrast, $\Phi_F^{(2)}(\omega)$ should display features near the characteristic frequencies of the system. In particular, for overdamped systems that we consider here such features occur around $\omega=0$. We note that $\Phi_F^{(1)}$ may also display features at the system eigenfrequency, since the integrand in Eq.~(\ref{eq:Phi_1_general}) depends on $\omega_F$.

A convenient way to calculate the fluctuating susceptibilities $\chi_{1,2}$ is based on solving dynamical equations of motion of the system. For example, for an overdamped Brownian particle described by the Langevin equation $\dot q=-U'(q)+ f(t) + F\cos\omega_Ft$ with nonlinear potential (\ref{eq:nonlinear_potential}), one can proceed by rewriting this equation in the integral form,
\begin{eqnarray}
\label{eq:integral_equation}
q(t)=&&\int_{-\infty}^t dt'e^{-\kappa(t-t')}\exp\left\{-\int_{t'}^tdt''\left[\beta q(t'')+\gamma q^2(t'')\right]\right\}\nonumber\\
&&\times\left[F\cos\omega_Ft' + f(t')\right].
\end{eqnarray}
For small $f$ and $F$, one can then expand the $q$-dependent exponential in the right-hand side and use successive approximations in $F$ and $f$. The fluctuating susceptibility $\chi_1$ is given by linear in $F$ terms, whereas $\chi_2$ is given by the terms quadratic in $F$.  The advantageous feature of this method is that it is not limited to white noise. However, the method becomes impractical if the noise intensity is not weak, and even for weak noise it becomes  cumbersome if one goes to high-order terms in the noise intensity.

We have checked that the calculation based on Eq.~(\ref{eq:integral_equation}) gives the same result for the driving-induced part of the power spectrum $\Phi_F(\omega)$ as the method of moments. We have also found that, in the second order in the noise intensity $D$, the term $\gamma q^4/4$ in $U(q)$ leads to the onset of a peak in $\Phi_F(\omega)$ at $\omega_F$.

\subsection{Fluctuating susceptibility of a threshold detector}
\label{subsec:suscept_threshold}

Fluctuating linear susceptibility has a particularly simple form for a threshold detector. By linearizing  in $F(t)$ expression (\ref{eq:threshold_model}) for the output of the detector, we obtain from the definition of the susceptibility (\ref{eq:susceptibility})
\begin{equation}
\label{eq:linear_fluct_suscept_threshold}
\chi(t,t') = 2\delta (t-t'-0)\delta\bigl(\xi(t)-\eta\bigr),
\end{equation}
where $\eta$ is the threshold and $\xi(t)$ is the noise. Zero in $\delta (t-t'-0)$ reflects causality: the detector output $q(t)$ is determined by the value of the driving just before the observation time; the very $\delta$-function indicates that the effect of the driving is not accumulated over time, the response is instantaneous (but causal).

The standard linear susceptibility $\chi(\omega)$ is given by expression 
\[\chi(\omega) = \int_0^\infty dt e^{i\omega t}\langle \chi(t,0)\rangle.\]
From (\ref{eq:linear_fluct_suscept_threshold}), $\chi(\omega)=2\rho^{\xi}_{\rm st}(\eta)$. where $\rho^{\xi}_{\rm st}(\eta)$ is the stationary probability density of the noise $\xi(t)$, cf. Eq.~(\ref{eq:susceptibility_threshold}). It applies for an arbitrary noise $\xi(t)$, not just for the exponentially correlated noise considered in Sec.~\ref{sec:threshold}.

Similarly, the fluctuating nonlinear susceptibility of the detector is
\begin{equation}
\label{eq:nonlinear_susc_threshold}
\chi_2(t,t',t'')= -\delta(t-t'-0) \delta(t-t''-0)\partial_\eta \delta\bigl(\xi(t)-\eta\bigr).
\end{equation}
Substituting Eqs.~(\ref{eq:linear_fluct_suscept_threshold}) and (\ref{eq:nonlinear_susc_threshold}) into the general expressions for the power spectrum in terms of fluctuating susceptibilities, Eqs.~(\ref{eq:Phi_1_general}) and (\ref{eq:Phi_2_general}), we obtain the power spectrum in the same form as what follows from Eq.~(\ref{eq:threshold_correlator}).



\section*{References}

\providecommand{\newblock}{}

\end{document}